\documentclass[proceedings]{JHEP} 

\newcommand{\be}[3]{\begin{equation}  \label{#1#2#3}}     
\newcommand{\ee}{ \end{equation}}
\newcommand{\ba}{\begin{array}}
\newcommand{\ea}{\end{array}}
\newcommand{\bea}{\begin{eqnarray}}
\newcommand{\eea}{\end{eqnarray}}
\newcommand{\NP}[3]{{\em Nucl. Phys.}{ \bf B#1#2#3}}

\def\ve{\varepsilon}
\newcommand{\ft}[2]{{\textstyle\frac{#1}{#2}}}

\def\beq{\begin{equation}}
\def\eeq{\end{equation}}
\def\beqa{\begin{eqnarray}}
\def\eeqa{\end{eqnarray}}

\renewcommand{\d}{\delta}
\newcommand{\pa}{\partial}
\newcommand{\g}{\gamma}

\newcommand{\e}{\epsilon}

\renewcommand{\L}{\Lambda}

\newcommand{\m}{\mu}

\newcommand{\n}{\nu}

\renewcommand{\o}{\omega}
\renewcommand{\O}{\Omega}

\conference{Quantum Aspects of Gauge Theories, Supersymmetry and Unification,
Paris 1999}

\title{Supersymmetric black hole solutions with $R^2$-interactions
\thanks{SPIN-00/10, ITP-UU-00/07, 
{\tt hep-th/0003157}}}

\author{G. L. Cardoso$^1$\thanks{Talk presented by G. L. C.}\,,
B. de Wit$^{1,2}$, J. K\"appeli$^1$ and T. Mohaupt$^3$
\\
$^1$Spinoza Institute, Utrecht University, 3584 CE Utrecht, The Netherlands\\

$^2$Institute for Theoretical Physics, Utrecht University,
3508 TA Utrecht, The Netherlands \\

$^3$Martin-Luther-Universit\"at Halle-Wittenberg, 
Fachbereich Physik,
D-06099 Halle, Germany \\
        E-mail: 
{\rm cardoso,bdewit,kaeppeli@phys.uu.nl, 
mohaupt@hera1.physik.uni-halle.de}

}

\abstract{We present a class of 
static supersymmetric multi-center black hole solutions
arising in four-dimensional 
$N=2$ supergravity theories with terms quadratic
in the Weyl tensor.  We also comment on possible 
corrections to the metric on the moduli space
of these black holes solutions. }

\begin{document}

\section{Introduction}

In this note we briefly describe a class of 
extremal static multi-center black hole solutions 
arising in four-dimensional 
$N=2$ supergravity theories with terms quadratic
in the Weyl tensor.  
These configurations preserve $N=1$ supersymmetry.  They 
are determined in terms of harmonic functions associated with
the electric and magnetic charges carried by the black holes.
We refer to an upcoming publication \cite{CDWKM} for a
detailed description of the construction of these solutions.


\section{Supersymmetry transformation \\ rules}
\setcounter{equation}{0}

The $N=2$ supergravity theories that we consider are based on vector
multiplets 
and hypermultiplets 
coupled to the supergravity fields and
contain the standard Einstein-Hilbert action as well as terms
quadratic in the Riemann tensor. To describe such 
theories in a transparent way we make use of the superconformal
multiplet calculus \cite{DWVHVPL}, 
which incorporates the gauge symmetries of the
$N=2$ superconformal algebra. The corresponding high  
degree of symmetry allows for the use of relatively small    
field representations. One is the Weyl multiplet, whose fields 
comprise the gauge fields corresponding to the superconformal 
symmetries and a few auxiliary fields. The other are abelian
vector multiplets 
and hypermultiplets, 
as well as a general chiral
supermultiplet.  The latter will be treated as independent in initial stages
of the analysis but at the end will be expressed in terms of the
fields of the Weyl multiplet.  Some of the additional (matter)
multiplets will provide 
compensating fields which are necessary in order that the 
superconformal action
becomes gauge equivalent to a Poincar\'e supergravi\-ty theory. The
compensating fields bridge the deficit in degrees of freedom
between the Weyl multiplet and the Poincar\'e supergravity multiplet.  
For instance, the graviphoton, represented by
an abelian vector field in the Poincar\'e supergravity multiplet,
is provided by an $N=2$ superconformal vector multiplet. 

It is possible to 
analyze the conditions for residual $N=1$ supersymmetry
directly in this superconformal setting, postponing a transition to
Poincar\'e supergravity till the end. This implies in particular that 
our intermediate results are subject to local 
scale transformations. Only towards the end we will convert to 
expressions that are scale invariant.  We will use this strate\-gy in the
following in order to construct black hole solutions with residual
$N=1$ supersymmetry.  This is exactly the same strategy we employed when we 
determined $N=2$ supersymme\-tric backgrounds in the presence
of $R^2$-inter\-ac\-tions \cite{CDWM}.

The superconformal algebra contains 
general-coordinate, local Lorentz, dilatation,
special conformal, chiral SU(2) and U(1), supersymmetry 
($Q$) and special supersymmetry ($S$) transformations. The gauge 
fields associated with  
general-coordinate transformations ($e_\m^a$), dilatations ($b_\m$), 
chiral symmetry (${\cal V}^{\;i}_{\m\, j}, A_\m$) 
and $Q$-supersymme\-try ($\psi_\m^i$), are 
realized by independent fields. The 
remaining gauge fields of Lorentz ($\o^{ab}_\m$), special 
conformal ($f^a_\m$) and $S$-supersymmetry transformations
($\phi_\m^i$) are dependent fields.  They are composite objects, which 
depend in a complicated way on the independent fields \cite{DWVHVPL}. 
The corresponding curvatures and covariant fields are contained
in a tensor chiral multiplet, which comprises $24+24$ off-shell 
degrees of freedom; 
in addition to the independent superconformal gauge fields it 
contains three auxiliary fields: a Majorana spinor doublet  
$\chi^i$, a scalar $D$ and a self\-dual Lorentz tensor $T_{abij}$ (where $i,
j,\ldots$ are chiral SU(2) spinor indices).
We summarize the transformation rules for some of
the independent fields of the Weyl multiplet
under $Q$- and $S$-super\-symmetry and under 
special conformal transformations, with parameters $\e^i$, $\eta^i$ and
$\L^a_{\rm K}$, respectively,
\bea \d e_\mu{}^a &=&
     \bar{\e}^i\g^a\psi_{\mu i}+{\rm h.c.}\,, \nonumber\\
      \d\psi_\mu^i &=& 2{\cal D}_\mu\e^i
      -\ft18  T^{ab\,ij}\g_{ab}\,\g_\mu\e_j
      -\g_\mu\eta^i \,,\nonumber\\
      \d b_\mu &=&
      \ft12\bar{\e}^i\phi_{\mu i}
      -\ft34\bar{\e}^i\g_\mu\chi_i
      -\ft12\bar{\eta}^i\psi_{\mu i}+{\rm h.c.} \nonumber\\
      &&+\L_{\rm K}^a\,e_\m^a  
      \,,\\   
      \d A_\mu &=& \ft{1}{2}i \bar{\e}^i\phi_{\mu i}
      +\ft{3}{4}i\bar{\e}^i\g_\mu\chi_i
      +\ft{1}{2}i \bar{\eta}^i\psi_{\mu i}+{\rm h.c.}\,, \nonumber\\
      \d T_{ab}^{ ij} &=&
      8 \bar{\e}^{[ i} {R}(Q)_{ab}^{j]}\,, \nonumber\\
      \d\chi^i &=& -\ft1{12}\g_{ab} D\!\llap/\, T^{ab\,ij}\e_j
      +\ft{1}{6} {R}({\cal V})_{ab}{}^{\!i}_{\;j}\,\g^{ab}\e^j \nonumber\\
      &&-\ft{1}{3}i {R}({A})_{ab} \g^{ab} \e^i 
      +D\,\e^i
      +\ft1{12} T^{ij}_{ab}\g^{ab}\eta_j \,, \nonumber
 \label{transfo4}
\eea
where ${\cal D}_\mu$ are derivatives covariant with respect to
Lorentz, dilatational, U(1) and SU(2) transformations,
whereas $D_\mu$ are derivatives covariant with respect to {\it all}
super\-conformal transfor\-ma\-tions.
The quantities ${R}(Q)^i_{\m\n}$,  
${R}(A)_{\m\n}$ and ${R}({\cal V})_{\m\n}{}^{\!i}_{\;j}$ 
are supercovariant curvatures 
related to $Q$-supersymme\-try, U(1) and SU(2) transformations.
We suppress
terms of higher order in the fermions through\-out this paper, 
as we will be dealing with
a bosonic background.

Let us now turn to the abelian vector multiplets, labelled by an 
index $I= 0,1,\ldots,n$. For each value of the index $I$,
there are $8+8$ off-shell degrees of freedom, residing in a 
complex scalar $X^{I}$, a doublet of chiral fermions 
$\Omega_i^{\,I}$, a vector gauge field $W_\mu^{\,I}$,
and a real SU(2) triplet of scalars $Y_{ij}^{\,I}$. 
Under $Q$- and $S$-supersymme\-try the fields $X^I$ and $\Omega_i^{\,I}$
 transform as follows:
\bea \d X^{I} &=& \bar{\e}^i\Omega_i^{\,I} \,,\nonumber\\
     \d\Omega_i^{\,I} &=& 2 D\!\llap/\, X^{I}\e_i
     +\ft12 \ve_{ij} (F_{\mu\nu}^{-I}
    -\ft14 \ve_{kl} T^{kl}_{\mu\nu}\bar{X}^{I})
    \g^{\m \n} \e^j \nonumber\\
     &&+Y_{ij}^{\,I}\e^j 
     +2X^{I}\eta_i\,,
\label{vrules}\eea
where the quantity $F_{\mu\nu}^{-I}$ denotes the anti-selfdual part of the
abelian field 
strength $F_{\mu\nu}^I=  2\pa_{[\mu}W_{\nu]}^I $.

The covariant quantities of the vector multiplet constitute a 
reduced chiral multiplet. A general chiral multiplet
comprises $16+16$ off-shell degrees of freedom
and carries an arbitrary Weyl weight $w$ (corresponding to the
Weyl weight of its lowest component). 
The covariant quantities of the Weyl multiplet also
constitute a reduced chiral multiplet, denoted by $W^{abij}$,
whose lowest-$\theta$ component is the tensor $T^{abij}$.
{}From this multiplet one may form a scalar (unreduced) chiral
multiplet $W^2 = [W^{abij}\,\varepsilon_{ij}]^2$ which has Weyl and
chiral weights $w=2$ and $c=-2$, respectively \cite{BDRDW}.  

In the following, we will also allow for the presence 
of an arbitrary chiral background
superfield \cite{deWit}, whose component fields will be indicated with a
caret.  
We denote its bosonic component fields by ${\hat A}$, ${\hat
B}_{ij}$,  ${\hat F}_{ab}^-$ and by ${\hat C}$.  Here ${\hat A}$ and 
${\hat C}$ denote complex scalar fields, appearing at the 
$\theta^0$- 
and $\theta^4$-level of the chiral background superfield, respectively,
while the symmetric complex SU(2) tensor ${\hat B}_{ij}$ and the
anti-selfdual Lorentz tensor ${\hat F}_{ab}^-$ reside at the 
$\theta^2$-level.  The fermion fields at level $\theta$ and 
$\theta^3$ are denoted by $\hat\Psi_i$ and $\hat\Lambda_i$. 
Under 
$Q$- and $S$-supersymmetry $\hat A$ and $\Psi_i$
transform as
\beqa
\d {\hat A} &=& \bar{\e}^i {\hat \Psi}_i \,,\nonumber\\
    \d {\hat \Psi}_i
 &=& 2 D\!\llap/\, {\hat A} \e_i
     +\ft12 \ve_{ij} {\hat F}_{ab} \g^{ab} \e^j
     +{\hat B}_{ij} \e^j \nonumber\\
     &&+2 w {\hat A} \eta_i\,,
\eeqa
where $w$ denotes the Weyl weight of the background superfield.
Eventually this multiplet will be identified with $W^2$ in
order to generate the $R^2$-terms in the action.
This identification implies the following
relations \cite{BDRDW}, which we will need in due time,
\bea
\hat A   &=& (\varepsilon_{ij}\,T^{ij}_{ab})^2\,,\nonumber \\
\hat \Psi_i &=& 16\, \varepsilon_{ij}R(Q)^j_{ab} \,T^{klab} \,
\varepsilon_{kl} \,,\nonumber\\  
\hat B_{ij}  &=& -16 \,\varepsilon_{k(i}R({\cal V})^k{}_{j)ab} \,
T^{lmab}\,\varepsilon_{lm} \nonumber\\
&&-64 \,\varepsilon_{ik}\varepsilon_{jl} 
\bar R(Q)^k_{ab} R(Q)^{l\,ab}   \,,\nonumber\\
\hat F^{-ab}  &=& -16 \,{\cal R}(M)_{cd}{}^{\!ab} \,
T^{klcd}\,\varepsilon_{kl}  \nonumber\\
&&-16 \,\varepsilon_{ij}\, \bar R(Q)^i_{cd} 
\gamma^{ab} R(Q)^{j\,cd}  \,,\nonumber\\
\hat \Lambda_i &=&32\, \varepsilon_{ij} \,\g_{ab} R(Q)_{cd}^j\, 
{\cal R}(M)_{cd}{}^{\!ab} \nonumber\\
&&+16\,({\cal R}(S)_{ab\,i} +3 \g_{[a} D_{b]}  \chi_i) \, 
T^{klab}\, \varepsilon_{kl} \nonumber\\
&& -64\, R({\cal V})_{ab}{}^{\!k}{}_i \,\varepsilon_{kl}\,R(Q)^l_{ab}
\,,\nonumber\\ 
\hat C &=&  64\, {\cal R}(M)^-_{cd}{}^{\!ab}\, {\cal 
R}(M)^-_{cd}{}^{\!ab}  \nonumber\\
&&+ 32\, R({\cal V})^-_{ab}{}^{\!k}{}_l^{~} \, 
R({\cal V})^-_{ab}{}^{\!l}{}_k^{~} \nonumber \\
&& - 32\, T^{ij\,ab} \, D_a \,D^cT_{cb\,ij} \nonumber\\
&&+128 \, \bar {\cal 
R}(S)^{ab}_i \,R(Q)_{ab}^i  \nonumber\\
&&+384 \,\bar R(Q)^{ab\,i} 
\gamma_aD_b\chi_i   \,.   
\label{w2}
\eea
We refer to \cite{CDWKM} for a precise definition of the various 
curvature tensors. The derivatives $D_a$ are superconformally covariant.

In the presence of a chiral background superfield, the coupling
of the abelian vector multiplets to the Weyl multiplet is encoded in a
function $F(X^I, {\hat A})$, which is holomorphic and
homogenous of degree two,
\beqa
&&X^I \, F_I + w {\hat A} \, 
F_{\hat A} = 2 F \;, \nonumber\\
&&F_I = \partial_{X^I} F \;,\quad
F_{\hat A} = \partial_{\hat A} F \;\;\;.
\label{homogeneity}
\eeqa
The field equations of the vector multiplets are subject to 
equivalence transformations correspon\-ding to electric-magnetic 
duality, which will not involve the fields of the Weyl multiplet
and of the chiral background. 
As is well-known, two complex $(2n+2)$-component vectors can be 
defined which transform linearly under the SP$(2n+2;{\bf R})$ 
duality group, namely
\beqa
\pmatrix{X^I\cr\noalign{\vskip1mm} F_I(X,{\hat A})} \; \quad 
{\mbox{ and}} \quad
\pmatrix{F_{\m\n}^{+I}\cr\noalign{\vskip1mm} G^+_{\m\n\,I}} \;.
\label{seca} 
\eeqa
The first vector has weights $w=1$ and $c=-1$, whereas the second 
one has zero Weyl and chiral weights. 
The field strengths $G^\pm_{\m\n\,I}$ are defined as follows:
\beqa
&&G^+_{\mu\nu I}=\bar F_{IJ}\,
F^{+J}_{\mu\nu} + {\cal O}_{\mu\nu 
I}^+\,, \nonumber\\
&&G^-_{\mu\nu I}= F_{IJ}F^{-J}_{\mu\nu} + 
{\cal O}_{\mu\nu I}^- \,, 
\label{defG}
\eeqa
where 
\bea
{\cal O}_{\mu\nu I}^+   =\ft14 (F_I-\bar F_{IJ}X^J )\,T_{\mu\nu 
ij}\varepsilon^{ij} +\hat F^+_{\mu\nu} \,\bar  
F_{I{\hat A}} \;. \nonumber\\
\label{NO} 
\eea
They appear in the field equations of the 
vector fields. 
Eventually we will solve the 
Bianchi identities, ${\cal D}^{\m} ( F^- - F^+)_{\m\n}^I = 0$, and the
field equations,  ${\cal D}^{\m} (G^- - G^+)_{\m\n}^I =0$, 
for a given configuration of magnetic and electric charges in a 
static spacetime geometry 
with the chiral background turned on.
These charges, which will be denoted by
$(p^I,q_I)$, 
comprise a symplectic vector.

Next, let us introduce a particular spinor that transforms
inhomogenously under $S$-super\-sym\-me\-try transformations.  This spinor
is given by 
\bea
&&\zeta_i^V \equiv - \Big(\Omega^I_i \,{\pa\over\pa X^I} + \hat\Psi_i \,
{\pa\over\pa \hat A}\Big) {\cal K} \\
&&=-i \,{\rm e}^{\cal K}\Big[  (\bar F_I - \bar 
X^JF_{IJ}) \Omega_i^I  -  \bar X^I F_{I{\hat A}} \, 
\hat\Psi_i\Big]\,, \nonumber
\eea
where we introduced the symplectically 
covariant factor (with $w=2$ and $c=0$),
\beq
{\rm e}^{-{\cal K}} = i\Big[\bar X^I\, F_I(X,\hat A) - \bar F_I(\bar X,
\bar{\hat A}) \, X^I \Big] \,,
\label{kaehler}
\eeq
which resembles (but is not equal to) the K\"ahler potential in 
special geometry.
It can be shown, using the results contained in \cite{deWit}, 
that $\zeta_i^V$ transforms covariantly under symplectic  
reparametri\-za\-tions. 
Under $Q$- and $S$-supersymmetry $\zeta_i^V$ transforms as 
(ignoring higher-order fermionic terms)
\bea
\d\zeta_i^V &=&   {\rm e}^{\cal K}
{\cal D}\!\llap/\, {\rm e}^{-\cal K} \e_i  
 +2 i 
{\cal A}\!\llap/  \,\e_i  - \ft12 i 
\ve_{ij} \, {\cal F}_{\m\n}^- \, \g^{\m\n} \e^j \nonumber \\
&& + {\rm e}^{\cal K} \Big[(\bar F_I - \bar X^J \bar F_{IJ} ) N^{IK}
 F_{KA} \, \hat B_{ij}  \nonumber\\
&&- (\bar F_I - \bar X^J F_{IJ} ) N^{IK}\bar
F_{KA} \,\varepsilon_{ik} 
\varepsilon_{jl} \hat B^{kl} \Big] \, \e^j \nonumber\\
&& +2  \,\eta_i\,,
\label{susyzeta}
\eea
where
\beqa
{\cal A}_{\mu} &=& \ft{1}{2} \, e^{\cal K}\, \Big( {\bar X}^J 
\stackrel{\leftrightarrow}{\cal D}_\m F_J - {\bar F}_J 
\stackrel{\leftrightarrow}{\cal D}_\m  X^J \Big) \;\;, \nonumber\\
{\cal F}_{\m\n}^- &=& {\rm e}^{{\cal K}} \, 
 \left( \bar F_I \,F^{-I}_{\m\n} - 
\bar X^I \,G_{\m\n\,I}^- \right) \;\;.
\label{curlyf}
\eeqa
In arriving at (\ref{susyzeta}) we have used the field equations for 
the auxiliary fields $Y_{ij}^I$ \cite{deWit}, which is ne\-cessary for
$\d\zeta_i^V $ to take a symplectically covariant form.

We note that 
$\d\zeta_i^V $ is not the only
spinor that can be constructed which transform 
inhomogenously under $S$-supersymme\-try transformations.  Another 
such spinor, which we denote by $\zeta_i^H$, is 
constructed out of the hypermultiplet fermions \cite{CDWKM}.  It transforms
as follows under 
$Q$- and $S$-supersymmetry, 
\bea
\d\zeta^{\scriptscriptstyle\rm H}_i = \ft12
\chi^{-1} {\cal D}\!\llap/\, \chi \,\varepsilon_{ij}\, \e^j 
+ k\!\llap/\,_{ij}\,\e^j  
 + \varepsilon_{ij}\, \eta^j \,, 
\eea
where $\chi$ denotes the hyper-K\"ahler potential and where $k_{\m ij}$
denotes a quantity 
that is symmetric in
$i,j$ \cite{CDWKM}
, but 
whose explicit form is not important here.

Since the $\zeta_i$ transform inhomogenously under 
$S$-supersymme\-try, they
can act as compensators for this symmetry.  This observation is relevant
when
constructing supersymmetric backgrounds, where one requires
(some of) 
the $Q$-supersymme\-try variations of the spinors (as well as
of derivatives of the spinors) to vanish modulo a
uniform $S$-transformation. This can conveniently be done by
considering $S$-invariant spinors, constructed by employing
$\zeta_i$. Relevant examples of such spinors are, for instance, 
$\O^I_i -
X^I\zeta_i^V$ and $\hat \Psi_i - w \hat A\, \zeta_i^V$.

\section{The ansatz}
\setcounter{equation}{0}

We seek to construct static multi-center black hole solutions 
with $N=1$ residual supersymmetry.
For the line element we make the ansatz 
\bea
ds^2 = - e^{2g(\vec{x})} dt^2 + e^{2f(\vec{x})} d \vec{x}^2 \;\;.
\label{line}
\eea
We impose the following restriction on the $Q$-super\-sym\-me\-try
transformation parameter $\epsilon_i$,
\bea
\epsilon_i = h \,  \varepsilon_{ij} \, \gamma_0 \, \epsilon^j  \;\;\;,
\label{ans}
\eea
where $h(\vec{x})$ denotes a phase factor of 
chiral weight $c=1$.  
The condition (\ref{ans}) is covariant with respect to SU(2) and
spatial rotations. The multi-cen\-ter solutions that we wish to
construct have the feature that, when the centers are made to
coincide, they lead to one-center solutions that are invariant with
respect to SU(2) and spatial rotations.  The latter satisfy condition
(\ref{ans}).

In addition to (\ref{ans}), we impose 
that 
${\cal A}_{\m} =0$
as well as ${\cal F}_{\m\n}^- =0$.  

We denote the magnetic and electric charges associated to each center by
$(p^I_A, q_{AI})$.  In the geometry (\ref{line}) the Bianchi identities
and field equations for the vector fields are solved by 
\beqa
F_{ti}^{-I} - F_{ti}^{+I} &=& - i \, e^{g-f} \partial_i H^I  
\;\;, \nonumber\\
G_{tiI}^{-} - G_{tiI}^{+} &=& - i \, e^{g-f} \partial_i H_I  
\;\;,
\label{gf}
\eeqa
where $H^I$ and $H_I$ denote harmonic functions given by
\bea
H^I &=& \sum_A \Big( h_A^I + \frac{p_A^I}{|\vec{x} - \vec{x}_A|} \Big) \,,\,
\nonumber\\
H_I &=& \sum_A \Big( h_{AI} + \frac{q_{AI}}{|\vec{x} - \vec{x}_A|} \Big) \;\;.
\eea
Here $ h_A^I $ and $ h_{AI} $ denote integration constants.
We now identify ${\hat A}$ with $(\ve_{ij} T^{abij})^2$ so that we are
dealing with black hole solutions in the presence of $R^2$-interactions.

\section{Static multi-center solutions}
\setcounter{equation}{0}

It will be convenient \cite{CDWM} to use rescaled variables
$Y^I = {\rm e}^{{\cal K}/2} {\bar \Sigma} X^I$ and 
$\Upsilon = {\rm e}^{\cal K} {\bar \Sigma}^2 {\hat A}$. Here $\bar \Sigma$
is taken to have weights $c=1$ and $w=0$ so that $Y^I$ and $\Upsilon$  
have vanishing
chiral and Weyl weights.  Then, from (\ref{kaehler}) and from 
(\ref{curlyf}), we obtain
\bea
|\Sigma|^2 &=& 
i\Big[\bar Y^I\, F_I(Y,\Upsilon) - \bar F_I(\bar Y,
\bar{\Upsilon}) \, Y^I \Big] \;\;, \nonumber\\
A_{\mu} &=& \ft{i}{2} \partial_{\m} \log \ft{\Sigma}{\bar \Sigma}
- {\cal A}^Y_{\m} \;\;,\\
{\cal A}^Y_{\m} &=& \ft{1}{2} \, \ft{1}{|\Sigma|^2}\, \Big( {\bar Y}^J 
\stackrel{\leftrightarrow}{\pa}_\m F_J - {\bar F}_J 
\stackrel{\leftrightarrow}{\pa}_\m  Y^J \Big) \;\;. \nonumber
\eea
Using (\ref{line}), we find \cite{CDWKM} that the vanishing of the 
$Q$-supersymmetry variation (subject to (\ref{ans})) of the various
$S$-invariant spinors 
yields a number of restrictions
on the $N=1$ background, as follows.  In a Poincar\'e frame (where
$b_{\m}=0$ and ${\cal K} = {\rm const.}$) we find that 
\bea
e^{-2g} &=& e^{2f} = {\rm e}^{{\cal K}}|\Sigma|^2 
\;\;,\;\; 
\nonumber\\
{\rm e}^{{\cal K}/2} {\bar \Sigma}
T_{ti}^- &=& 4 \, \partial_i f
\;\;,\;\; \Upsilon = - 64 (\partial_i f)^2 \;\;,
\nonumber\\
F_{ti}^I &=& - \partial_i \Big[ e^{-2f} (Y^I + {\bar Y}^I) \Big]
\;\;,\nonumber\\
 {\cal A}_{\m}^Y &=& 0 \;\;.
\label{n1bg}
\eea
The symplectic vector $(Y^I, F_I (Y, \Upsilon))$ is determined in terms
of the symplectic vector $(H^I,H_I)$ as follows:
\bea
\partial_i 
\pmatrix{Y^I - {\bar Y}^I \cr\noalign{\vskip1mm} 
F_I(Y,\Upsilon) - {\bar F}_I ({\bar Y}, {\bar \Upsilon})} 
= i\, \partial_i 
\pmatrix{H^I \cr\noalign{\vskip1mm} H_I} \;. \nonumber\\
\eea
Thus, we see that as one approaches the indivi\-dual centers 
($|\vec{x} - \vec{x}_A| \rightarrow 0$) 
the scalar fields $Y^I$ are entirely 
determined in terms of the charges associated
to these indivi\-dual centers.  This behaviour, namely that the values of the
$Y^I$ near the centers are independent of 
the constants
$h^I_A$ and $h_{AI}$ which determine 
the values of the $Y^I$ 
far away from the centers, is the same that has been observed without
$R^2$-interactions \cite{FKS}.

In addition, we find that $h = {\bar \Sigma}/|\Sigma|$, whereas
the field $D$ is determined to be $D = - \ft{1}{3} R$.
The SU(2) curvature $R({\cal V})_{\m\n}\,^i\,_{ j}$, and hence 
${\hat B}_{ij}$,
vanishes.  Since ${\cal A}^Y_{\m}=0$, also the U(1) curvature $R(A)_{\m\n}$
vanishes.   

The solution given above describes a static mul\-ti-cen\-ter black ho\-le 
with residual $N=1$ supersymmetry in the presence of $R^2$-interactions.
It approaches flat Minkowski spacetime at spatial infinity.
Setting
${\rm e}^{\cal K} |\Sigma|^2_{\infty} = 1$ 
expresses its ADM mass as
$M_{\rm ADM} = {\rm e}^{\cal K} \sum_A (p_A^I F_I (Y_{\infty})
- q_{AI} Y^I_{\infty})$.  

In the case of one center, the solution interpolates
between two $N=2$ supersymmetric vacua \cite{Gibbons}: 
flat spacetime at spatial infinity
and Bertotti-Robinson space\-time at the hori\-zon.  When switching off
$R^2$-interactions this solution agrees with the one constructed in
\cite{BLS}.
Its macroscopic
entropy is given by \cite{CDWM}
\beqa
{\cal  S} = \pi r^2  \Big[ |\Sigma|^2 +4 \, {\rm Im} \Big( 
\Upsilon \,F_{\Upsilon}
(Y,\Upsilon) \Big)
\Big]\Big|_{r=0} .
\label{entropiay}
\eeqa

\section{Outlook}
\setcounter{equation}{0}

The static multi-center solution (\ref{n1bg}) can now be used as the
starting point for computing the me\-tric on the moduli space 
of four-dimensional BPS black holes in the presence of $R^2$-interactions.
In the absence of $R^2$-interactions, it was found \cite{Stro,Papa}
that the moduli metric of electrically
charged BPS black holes is determined in terms of a mod\-uli potential $\mu$
given by $\mu = \int d^3x\,  e^{4f}$.  It was furthermore established
\cite{Stro,Papa}
that for small black hole separations the associated one-dimensional
Lagrangian describing the slow-motion of these BPS black holes 
exhibits an enhanced superconformal symmetry.  It 
was suggested 
\cite{Stro} that it should be possible to reproduce the macroscopic
entropy of BPS black holes by performing a state counting in this
superconformal quantum mecha\-nics model.  This would imply that the
degene\-ra\-cy of states of such a model is encoded in the moduli
potential $\mu$.  In view of the formula (\ref{entropiay}) for the
macroscopic entropy one thus expects
that $\m$ will 
receive corrections steming from $R^2$-inter\-ac\-tions.  This is indeed
likely to be the case, since the one-dimensional Lagrangian describing
the slow-motion of the black holes will now be derived from a 
four-dimensio\-nal
action containing $R^2$-interactions.   Schematically, 
since the four-dimen\-sio\-nal Einstein-Maxwell action gi\-ves rise 
to a moduli
potential $\mu = \int d^3x \, e^{ 4f}$, the
term 
${\rm Im} F_{{\hat A}} \, R \, |T|^2$, which appears in the 
four-dimen\-sio\-nal
Lagrangian 
with $R^2$-interactions, suggests
a correction to the moduli potential
of the form 
$\int d^3x \, e^{2f}  {\rm Im} (\Upsilon
F_{\Upsilon} ) $.  In view of (\ref{entropiay}), this suggests that
in the presence of $R^2$-terms the moduli potential $\m$ will be given by
$\mu = \int d^3x \, e^{2f} 
[e^{2f} +4 \, {\rm Im} (
\Upsilon \,F_{\Upsilon}
(Y,\Upsilon) ) ]$.  This feature is currently under investigation
\cite{CDWKM2}.



\section*{Acknowledgements}

This work was supported in part by the European 
Commission TMR programme ERBFMRX-CT96-0045.


\end{document}